\newcommand{\lsim}{\stackrel{\mbox{\raisebox{-0.1ex}{\scriptsize $<$}}}
{\mbox{\raisebox{-0.5ex}{\scriptsize $\sim$}}}}

\documentstyle[12pt]{article}
\newlength{\myleftmargin}
\newlength{\paperwidth}
\begin{document}

\thispagestyle{empty}
\begin{flushright}
KOBE-FHD-95-07  \vspace{2ex} \\
November 1995 \vspace{2ex} \\
\end{flushright}

\begin{center}
\renewcommand{\thefootnote}{\fnsymbol{footnote}}
 {\large \bf Moduli effects on neutrino oscillations}\\

(published in Phys.Rev.D {\bf 54}, 1204-1211, 1996)\\
\vspace{4em}
K. Kobayakawa${}^{1}$, 
Y. Sato${}^{2}$\footnote{satou@jet.earth.s.kobe-u.ac.jp},  
and S. Tanaka${}^{3}$ 
\footnote{present address: Department of Physics, 
Hyogo University of Education, Yashiro-cho, Hyogo, 
673-14, Japan}\\
\vspace{3em}
${}^{1}$
        {\it Fukui University of Technology, }\\
        {\it Gakuen, Fukui 910, Japan}\\

\vspace{2em}
${}^{2}${\it Graduate School of Science and Technology, }\\
        {\it Kobe University, Nada, Kobe 657, Japan}\\
\vspace{2em}
${}^{3}$
        {\it Division of Natural Environment and High Energy Physics, }\\
        {\it Faculty of Human Development, }\\
        {\it Kobe University, Nada,Kobe 657, Japan}\\
\end{center}

\begin{center}
\begin{abstract}
\baselineskip=20pt
We point out the possibility of detecting low-energy signals of moduli in 
superstring theory through neutrino oscillations. The idea is based on the 
characteristics that the couplings of moduli are different from matter to 
matter. We estimate the oscillation probability both in the base line and 
solar neutrino oscillations. In both cases, when there is at least one 
modulus of which the mass is less than or equal to $10^{-19}$GeV, the 
interaction of the modulus significantly changes the 
conversion probability from one neutrino flavor to another.
\end{abstract}
\end{center}

\clearpage
\pagestyle{plain}
\setcounter{page}{1}
\baselineskip=16pt

{\large \bf 1.~Introduction}

\vspace{1em}
Recent data from the CERN $e^{+}$ $e^{-}$ collider LEP \cite{LEP}
suggest the evidence of grand unified theories (GUT's)
such as SU(5), SO(10), flipped SU(5) and so on. Furthermore, 
the data fit better if supersymmetry is included. 
On the theoretical side, to solve the gauge hierarchy problem 
the idea of supersymmetry is very persuasive. 
However, a SUSY GUT does not contain the interaction of gravity. 
At present it is conceived that superstring theory alone 
may include all interactions consistently in the theory. 
Phenomenologically the heterotic superstring theory 
\cite{Gross} is the most attractive. 
There are several ways of compactification, and 
after that very many vacua are prodced \cite{Green}. 
They are parametrized, in general, by moduli \cite{Ferrara}
which are singlet superfields under the 
gauge group of the standard model, 
$\rm{SU(3)_{C}}\times \rm{SU(2)_{L}}\times \rm{U(1)_{Y}}$. For example,
some of them describe the size and shape of compactified space.
Although their vacuum expectation values (VEV's) are supposed to be of 
the order of Planck scale, masses of moduli are not known. 
Their interactions with matter are also 
model dependent. Even the number of moduli depends on the structure 
of the vacuum under consideration. The number of K\"ahler structure 
moduli, $T_{i}$, is given by the Hodge number $h^{(1,1)}$ and that 
of complex structure moduli, $U_{m}$, is $h^{(2,1)}$. They are, in 
general, large numbers in (2,2) Calabi-Yau manifolds. 
In (0,2) orbifolds, however, the former is at most 9 
and the latter is at most 3
[there are other types of moduli, too: (0,2) untwisted moduli 
(Wilson lines) and twisted moduli]. 
In any case, there exist moduli. 
Both $T_{i}$ and $U_{m}$ behave similarly as particles.
Since moduli have a very important role in superstring theory, it is 
very helpful to detect the moduli.

In this paper\footnote[1]{This is a revised paper of the preprint 
KOBE-FHD-95-04, hep-ph/9504370.} 
we would like to point out that moduli may give low-energy 
signals which could be tested in the neutrino oscillation experiments 
without depending on a particular compactification scheme.
Moduli generically couple to ordinary matter with nonrenormalizable 
interactions. Such couplings are expressed effectively in 
the superpotential as (in the lowest dimension) 
\begin{equation}
P_{nonren} = \frac{c^{I}_{ijk}}{M_{\rm{S}}}\varphi_{i}\varphi_{j}
\varphi_{k}M_{I}, ~~~~~~~~~~~~~~~(I=1,2,3,\dots),
\end{equation}
where $\varphi_{i,j,k}$ are matter superfields, 
$M_{I}$ are moduli superfields and $M_{\rm{S}}$ is the string scale
($\sim 10^{18}\rm{GeV})$. $c_{ijk}$ may contain a product of VEV's of 
many scalar fields \cite{Hyokyo}.
Such terms at low energies induce 
Yukawa-type couplings between the ordinary matter 
and (real) scalar fields, or pseudoscalar fields i.e.
moduli:
\begin{eqnarray}
{\cal L}_{Y} &=& \frac{<H_{2}>}{M_{\rm{S}}}h_{ij}^{(\nu)}
\bar{\nu}^{i}_{R}{\nu}^{j}_{L}M_{I} + 
\frac{<H_{2}>}{M_{\rm{S}}}h_{ij}^{(u)}
\bar{u}_{R}^{i}u_{L}^{j}M_{I} \nonumber \\ 
&+& \frac{<H_{1}>}{M_{\rm{S}}}h_{ij}^{(d)}
\bar{d}_{R}^{i}d_{L}^{j}M_{I} + 
\frac{<H_{1}>}{M_{\rm{S}}}h_{ij}^{(\ell)}
\bar{\ell}_{R}^{i}\ell_{L}^{j}M_{I} + h.c.,
\label{eqn:Yukawa}
\end{eqnarray}
where $i$ and $j$ are generation indices ($i=1,2,3$) and $<H_{1,2}>$ 
are the vacuum expectation values of the Higgs doublets, 
and $\gamma$-matrices are dropped. 
While the dilaton $S$ interacts with ordinary matter universally 
\footnote[2]
{This is the case at the string tree level. At the loop level, this 
universality is lost. And so the dilaton may take part in the 
neutrino oscillation, too. }
like a graviton, moduli interact (or non-interact) with various coupling 
constants. Moduli interact with ordinary matter as a coherent attractive or 
repulsive force. 
Since the interaction strength is comparable to that of 
gravity force, this 
behaves as a kind of fifth force if the mass of the exchanged particle 
is small enough \cite{Cvetic,Macorra}. 
The potential of moduli is considered flat perturbatively to all orders. 
When spontaneous breaking of SUSY occurs, most or all moduli may get mass 
by nonperturbative effects. Therefore their masses are expected to be 
of the order of the gravitino mass. 
Namely, it would be as heavy as other scalar sparticles. 
But a few may have very tiny mass or massless after SUSY breaking.
There are several arguments which support it: 

(1) For real $M_{I}$ the moduli mass $\mu$ may be 
induced by radiative corrections 
($\mu\simeq 10^{-18} \rm{GeV}$) \cite{Cvetic}, 
or there may be a special cancellation in the mass equation. In 
Ref. \cite{Kelley}, it is estimated that $\mu$ 
can be about $m_{\frac{3}{2}}^{2}/{\rm Re}M_{I}$, where 
$m_{\frac{3}{2}}$ is the gravitino mass.

(2) For imaginary $M_{I}$, in Ref. \cite{Macorra} it was argued 
that $\mu$ can be $2\times10^{-24}\rm{GeV}$. 
However, in Ref. \cite{Kelley} it is said that they are massless. 
In Ref. \cite{Ibanez}, on the other hand, they are said to gain huge 
mass of the order of the SUSY-breaking scale.

We do not go into details of the models here and want to discuss 
model-independent way as much as possible. We regard a mass of a 
modulus (especially a tiny one) as a free parameter and its interaction 
strength as parameters $f_{ij}$, and explore the possibility 
of finding the effects of moduli in terrestrial experiments, 
not in cosmology.

Section 2 has two subsections. In section 2.1, 
taking the influence of moduli interaction into consideration, 
we obtain the oscillation probability. 
In section 2.2 we examine how the moduli interaction affects 
the planning experiments. In section 3, we estimate 
the moduli effect on the solar neutrino oscillation. 
In section 4, we argue the problematic points and 
mention a prospect of future experiments. 
\vspace{1em}

{\large \bf 2.~Moduli Effects} \\

{\large \bf 2.1.~Oscillation Probability} \\

In this section we deal with the accelerator experiments and derive 
the $\nu_{\mu}$-$\nu_{\tau}$ oscillation probability 
including the effect of moduli interaction.
We assume that there is, for simplicity, at least one modulus 
which interacts with $\nu_{\tau}$ 
and/or $\nu_{\mu}$ and {\it u} or {\it d} quark (or electron). 
For example, $h_{33}^{(\nu)}\neq 0$, $h_{11}^{(u)}\neq 0$, and 
others can be zero in Eq. (\ref{eqn:Yukawa}). 
Although the interaction strength is gravitational, it may be detectable 
in the neutrino oscillations 
when $\mu$ is very tiny. 
We take $\mu$ in the range of $10^{-22}-10^{-14}$GeV. 
 
We define the mass eigenstate as 
$(\nu_{2}^{m}, \nu_{3}^{m})$, and 
the flavor eigenstate as $(\nu_{\mu}, \nu_{\tau})$. 
The latter eigenstate is expressed by the former with a mixing angle 
$\theta$ as
\begin{equation}
\left(
\begin{array}{l}
\nu_{\mu} \\
\nu_{\tau}
\end{array}
\right)
= U
\left(
\begin{array}{l}
\nu_{2}^{m} \\
\nu_{3}^{m}
\end{array}
\right), ~~~
U = 
\left(
\begin{array}{cc}
\cos\theta      & \sin\theta\\
-\sin\theta     & \cos\theta
\end{array}
\right).
\label{eqn:mixing}
\end{equation}
The neutrino interaction with matter through moduli which is derived 
from Eq. (\ref{eqn:Yukawa}) can be replaced by the Yukawa potential 
as moduli interact coherently and we put its coupling constants as 
$f_{ij}G_{M}$. $G_{M}$ is the common coupling constant of the modulus 
so that the maximum value among $|f_{ij}|$ is unity. 
The Hamiltonian of the mass eigenstate is changed to 
\begin{equation}
H = \left(
\begin{array}{cc}
p+\frac{m_{2}^{2}}{2p}-f_{22}'\phi      & -f_{23}'\phi\\
-f_{32}'\phi                           & p+\frac{m_{3}^{2}}{2p}-f_{33}'\phi
\end{array}
\right),
\label{eqn:Hamiltonian}
\end{equation}
where $p$ is the momentum of a neutrino beam, and $m_{2}$ and $m_{3}$ 
are the masses of mass eigenstates. 
In Eq. (\ref{eqn:Hamiltonian}) $f_{ij}'\phi$ represent the potentials 
induced by moduli interaction and  
\begin{equation}
\left(
\begin{array}{cc}
f_{22}'  & f_{23}' \\
f_{32}'  & f_{33}' \\
\end{array}
\right)=
U^{-1}\left(
\begin{array}{cc}
f_{22}  & f_{23} \\
f_{32}  & f_{33} \\
\end{array}
\right)U.
\end{equation}
We can take $f_{23}=f_{32}$. Because of minus signs before 
$f_{ij}'\phi$, $\phi>0$ means that it is an attractive potential and 
$\phi<0$ means repulsive.
At least in orbifold models \cite{orbifold} either the diagonal ($f_{33}$ 
or $f_{22}$) or non-diagonal ($f_{23}$) part may be considered to 
vanish or to be very small. 
Let us consider the following simple two cases: (A) $\Delta f=1$ 
($\Delta f\equiv f_{33}-f_{22}$), $f_{23}=0$; (B) $\Delta f=0$, 
$f_{23}=1$. 

The flavor eigenstate obeys the Schr\"{o}dinger-like 
matrix equation,
\begin{equation}
i\frac{d}{dx}
\left(
\begin{array}{l}
\nu_{\mu} \\
\nu_{\tau}
\end{array}
\right)
= UHU^{-1}
\left(
\begin{array}{l}
\nu_{\mu} \\
\nu_{\tau}
\end{array}
\right).
\label{eqn:motion}
\end{equation}
It does not make any difference to the probability of the $\nu_{\mu}-
\nu_{\tau}$ transition if we subtract from $UHU^{-1}$ any multiple 
of the unit matrix. We choose the Hamiltonian matrix 
traceless for the sake of convenience: namely,
\begin{eqnarray}
i\frac{d}{dx}
\left(
\begin{array}{l}
\nu_{\mu} \\
\nu_{\tau}
\end{array}
\right)
=
\left(
\begin{array}{cc}
-a & b \\
b &  a
\end{array}
\right)
\left(
\begin{array}{l}
\nu_{\mu} \\
\nu_{\tau}
\end{array}
\right), 
\label{eqn:ab}
\end{eqnarray}
where
\begin{eqnarray}
a &\equiv& \frac{\Delta m^{2}}{4E}\cos2\theta- \frac{\Delta f}{2}\phi, 
\label{eqn:adef} \\
b &\equiv& \frac{\Delta m^{2}}{4E}\sin2\theta- f_{23}\phi, 
\label{eqn:bdef}
\end{eqnarray}
and $\Delta m^{2}\equiv m_{3}^{2}-m_{2}^{2}$. The momentum $p$ is 
replaced by the neutrino energy $E$ hereafter. 
Solving this, we obtain the oscillation probability 
\begin{eqnarray}
P(\nu_{\mu}\rightarrow\nu_{\tau})
&=& \sin^{2}2\theta_{M} \nonumber \\
&\times& \sin^{2} \left\{
\left[
\left(\frac{\Delta m^{2}}{4E}
\right)^{2} + 
\left(\frac{\Delta f'\phi}{2}
\right)^{2} 
- \frac{\Delta m^{2}}{4E}\Delta f' \phi
+ (f'_{23}\phi)^{2}
\right]^{\frac{1}{2}}L
\right\}.
\label{eqn:oscillation1}
\end{eqnarray}
It is rewritten as 
\begin{equation}
P(\nu_{\mu}\rightarrow\nu_{\tau})=
\frac{b^{2}}{a^{2}+b^{2}}\sin^{2}
(\sqrt {a^{2}+b^{2}} \cdot L),
\label{eqn:oscillation2}
\end{equation}
where $\theta_{M}=\theta + \zeta$ ($\zeta$ is the mixing angle from 
eigenstate of $H$ to mass eigenstate), $\tan 2\theta_{M}=b/a$, and so 
\begin{equation}
\sin^{2}2\theta_{M}=\frac{b^{2}}{a^{2}+b^{2}}.
\label{eqn:sinedef}
\end{equation}
$L$ is the distance between an accelerator and a detector. 
The first term inside the brackets in Eq. (\ref{eqn:oscillation1})
is due to the oscillation in the vacuum, and the last three terms 
are due to moduli interaction. 

Next we evaluate $\phi$ in base line neutrino experiments.
In a relativistic case $\phi$ is represented as the product 
of energy of a neutrino beam and the potential per unit mass 
due to moduli interaction with matter{\cite Gasperini}. 
For an attractive force we get 
\begin{eqnarray}
\phi &=& EV, \nonumber \\
V &=& G_{M} \frac{M}{r}\exp(-\mu r).
\label{eqn:exponential}
\end{eqnarray}
Here $M$ is the mass of the matter which interacts with the neutrino 
by interchanging moduli. $\phi$ changes its sign for a repulsive 
force. There may be a case that $\Delta f=-1$ and $f_{23}=0$. In this 
case the attractive force gives the same results as those of the 
repulsive one in case (A) when both coupling constants are 
equal to each other. So we will not discuss the case of $\Delta f=-1$ 
and $f_{23}=0$. 
To estimate $V$, we consider the following two cases.

(1) The contribution to $V$ from the whole Earth is added up. 
We assume the density $\rho$ to be constant. Then 
\begin{equation}
\phi_{\rm global}=
\frac{2\pi G_{M}\rho E}{\mu^{2}}\left\{2-\frac{R_{E}+\mu^{-1}}{z_{0}} 
[e^{-\mu(R_{E}-z_{0})}-e^{-\mu(R_{E}+z_{0})}]\right\}, 
\label{eqn:global}
\end{equation}
where $R_{E}$ denotes the radius of the Earth. 
$z_{0}$ is the average distance between the neutrino trajectory 
and the center of the Earth (see Fig.1). 
We can put $z_{0}\simeq R_{E}-\frac{L^{2}}{12R_{E}}$. 
Since in the planning base line experiments $L\ll 2R_{E}$ and the main 
contribution to $V$ comes from the parts near the neutrino trajectory, 
we put $\rho$ to be the density of the surface layer of the Earth: 
$\rho=\rho_{\rm sur}=2.76$[g$\cdot\rm{cm}^{-3}$].

(2) $\mu^{-1}$ is the scale of the region where moduli interaction 
is effective. Consequently, the sphere within the radius $\mu^{-1}$ 
is sufficient for the estimation of $V$. We obtain 
\begin{equation}
\phi_{\rm local} = \frac{4\pi G_{M}\rho_{sur}E}{\mu^{2}}
\left(1-\frac{2}{e}\right).
\label{eqn:local}
\end{equation}
The value of $\phi$ given by Eq. (\ref{eqn:global}) is almost tantamount 
to that of Eq. (\ref{eqn:local}) because of 
the exponential damping appearing in Eq. (\ref{eqn:exponential}).
Hence we use Eq. (\ref{eqn:global}) hereafter. 

As is obviously seen from Eq. (\ref{eqn:global}) or Eq. 
(\ref{eqn:local}) the potential $\phi$ is proportional to $\mu^{-2}E$, 
that is to say, the smaller $\mu$ is, the larger the effect of moduli 
is. The effect of the moduli is enhanced by $E^{2}$ relative to the 
vacuum oscillation part $\Delta m^{2}/4E$. In case (A) a very large 
$\phi$, i.e., $|\Delta f\phi/2|\gg \frac{\Delta m^{2}}{4E}\cos\theta$ 
and so $a^{2}\gg b^{2}$, leads to a very small $P(\nu_{\mu}\rightarrow
\nu_{\tau})$. On the other hand, in case (B) a large $\phi$ means that 
$b^{2}\gg a^{2}$ and the magnitude $\sin^{2}2\theta_{M}=b^{2}/(a^{2}+
b^{2})$ approaches 1. Then the oscillation length defined by $\ell=
\pi/\sqrt{a^{2}+b^{2}}$ is much smaller than the oscillation length 
in the vacuum, $\ell_{v}=4\pi E/\Delta m^{2}$. Therefore, for 
$L\simeq \ell_{v}$ the probability is averaged to be a half of the 
magnitude. Contrary to these, when $|\phi|\ll\frac{\Delta m^{2}}{4E}
\cos\theta$ (or $\sin\theta$), the effect cannot be seen. 

It is noted that a resonance similar to the solar neutrino oscillation 
occurs under a certain condition in case (A). When 
\begin{equation}
\frac{\Delta m^{2}}{4E}\cos2\theta=
\frac{\Delta f}{2}\phi,
\label{eqn:resonance}
\end{equation}
then $a=0$, and the magnitude is unity. Usually $\Delta m^{2}$ is 
considered to be positive, and so the resonance occurs in the attractive 
(repulsive) force for positive (negative) $\Delta f$. On the contrary, 
if $\frac{\Delta m^{2}}{4E}\sin 2\theta\simeq f_{23}\phi$ in case (B) 
with the attractive force, then $b\simeq 0$ and 
$P(\nu_{\mu}\rightarrow\nu_{\tau})$ is strongly supressed. 

\vspace{1em}
{\large \bf 2.2~Oscillations on Long and Short Base lines} \\

In this section we discuss long and short base line neutrino 
oscillations. In the planning experiments the muon neutrino 
($\nu_{\mu}$) beam with energy $E$ (of the order of 1 GeV to a few 
10 GeV) propagates along the trajectory. 

We now evaluate the oscillation probability. The force induced by the 
interaction of moduli with very tiny mass behaves like a fifth force, 
which many experiments have tested and given limitations to. 
Restrictions on the coupling constant $G_{5}$ as a function 
of the range $\lambda$ have been given. First fixing the value of 
$\mu$ where $\mu=\lambda^{-1}$, we take $G_{M}$ in Eq. 
(\ref{eqn:exponential}) at the maximum value of allowable $G_{5}$.
Denoting $\alpha=\frac{G_{M}}{G_{N}}$, where $G_{N}$ is the 
gravitational constant, we impose restrictions
for the attractive force from Ref. \cite{Nature}; 
for example, $(2.0\times10^{-22},3.0\times10^{-6})$, 
$(2.0\times10^{-20}, 1.6\times10^{-4})$, 
$(2.0\times10^{-18}, 5.0\times10^{-4})$, 
in terms of $(\mu [{\rm GeV}], \alpha)$ (see Table 1). 
Similarly, for the repulsive force the restrictions are found in 
Ref.\cite{Stacey}(see Table 3). 

We will comment on Eq. (\ref{eqn:oscillation1}) here. 
The quantity in the braces can be written as 
\begin{equation}
\left(\frac{\Delta m^{2}}{4E}
-\frac{\Delta f}{2}\phi\right)L, 
\label{eqn:brace}
\end{equation}
for $f_{23}=0$ and $\cos 2\theta =1$. 
In order to estimate the moduli effect roughly, we compare the value 
due to $\phi$ with the vacuum part in the following way. As is obvious, 
\begin{equation}
\frac{\Delta m^{2}}{4E}L=1.27
\frac{\left(\frac{\Delta m^{2}}{{\rm eV}^{2}}\right)}
{\left(\frac{E}{{\rm GeV}}\right)}
\left(\frac{L}{{\rm km}}\right),
\label{eqn:vacuum}
\end{equation}
For $\phi$, using Eq. (\ref{eqn:global}),  
\begin{eqnarray}
\frac{1}{2}\phi L &\simeq& \frac{\pi G_{M}\rho EL}{\mu^{2}}  
\nonumber \\
&=& 1.23 \left(\frac{\alpha}{10^{-4}}\right)
\left(\frac{\mu}{10^{-20}{\rm GeV}}\right)^{-2}
\left(\frac{E}{{\rm GeV}}\right)
\left(\frac{L}{{\rm km}}\right). 
\label{eqn:moduli}
\end{eqnarray}
When all physical quantities are the same in the denoted units, 
both values of Eqs. (\ref{eqn:vacuum}) and (\ref{eqn:moduli}) 
are almost the same and close to $\pi/2$. The above two equations 
are also useful to calculate $a$ and $b$ given by Eqs. (\ref{eqn:adef}) 
and (\ref{eqn:bdef}).

Let us consider two versions of $\Delta m^{2}$ and $\theta$. 
First, if $\nu_{\tau}$ 
is regarded as a candidate of dark matter, then $\Delta m^{2}$ is 
expected to be about $100{\rm eV}^{2}$ \cite{Harari} in which 
the mixing angle is supposed to be very small 
($\theta\simeq 1.0\times 10^{-2}$). 
Second, according to Kamiokande atmospheric neutrino data, 
$\Delta m^{2}\simeq 10^{-2}{\rm eV}^{2}$ \cite{Kamiokande} 
and the maximal mixing ($\theta\simeq\frac{\pi}{4})$ is suggested. 

The short bese line experiments, such as 
(i) CHORUS ($E=10$GeV, $L=0.8$km) \cite{CHORUS}, 
expect the former version. We take $\theta =1.0\times10^{-2}$ 
and $\Delta m^{2}=100\rm{eV}^{2}$ for this experiment. 
The long bese line experiments, such as 
(ii) KEK $\rightarrow$ Kamioka ($E=1.4$GeV, $L=250$km) \cite{KEK}, 
(iii) Fermilab $\rightarrow$ SOUDAN2 ($E=10$GeV, $L=800$km) \cite{FNAL}, 
expect the latter version. So we fix 
$\theta=\pi/4$ and $\Delta m^{2}=10^{-2}\rm{eV}^{2}$.

Our results are shown in Table 1-4. 
The first column shows the values of $\lambda$, the second does the 
values of $\mu$, and the third is assigned to $\alpha$'s in Table 1 
and 3. When the values of $\lambda$ are fixed, the probability can 
be calculated from Eq. (\ref{eqn:oscillation2}). Here we evaluate the 
following two quantities involved in the formula of the probability: 
\begin{eqnarray}
\sin^{2}2\theta_{M} &=& \frac{b^{2}}{a^{2}+b^{2}}, 
\label{eqn:sinedef} \\
\frac{\pi}{\ell} &=& \sqrt{a^{2}+b^{2}}.
\end{eqnarray}
$\sin^{2}2\theta_{M}$ represents the magnitude of probability. 
$\ell$ is the oscillation length. So $P(\nu_{\mu}\rightarrow
\nu_{\tau})=0$ when $L=\ell$. We list these values in each table. 
The results for the attractive force are listed in Tables 1 and 2, and 
those for the repulsive force in Tables 3 and 4.

Table 1 represents the estimation for the CHORUS experiment. In case 
(A), $\sin^{2}2\theta_{M}$ is reduced to the comparatively lower 
values in the whole range of $\mu$ because $b$ is small and constant. 
When $\mu$ is small, the probability changes rapidly on account of 
large values of $\pi\ell^{-1}$. When $\mu$ is larger than about 
$4\times 10^{-19}$GeV, no moduli effects can be seen:
$P(\nu_{\mu}\rightarrow\nu_{\tau})$ shows no difference from the 
oscillation in the vacuum. 
A particular value of $\mu$ causes a phenomenon like a resonance which 
gives the largest value to $\sin^{2}2\theta_{M}$. 
We will discuss this phenomenon in more detail later. 
In case (B), for a small value of $\mu$, $\sin^{2}2\theta_{M}$ is 
nearly unity, but $\ell$ is very small. 
Therefore, the probability is supposed to be averaged to one-half 
and this may be observable. When $\mu$ is heavier, $\sin^{2}2\theta_
{M}$ is smaller and neutrino oscillates more slowly to make little 
difference than that in the vacuum. In this case, however, an incident 
which we may call "antiresonance" occurs when $\mu$ takes the value 
such as $b$ vanishes; namely, the moduli effect cancels the oscillation 
in the vacuum. 

Our calculations on KEK and Fermilab-experiments are listed in Table 2 
for the attractive case. Here we take the angle $\theta=\frac{\pi}{4}$, 
and so the first cosine term in Eq. (\ref{eqn:adef}) is zero. As is 
seen in case (A) of KEK, only in a narrow range of $\mu, 10^{-19}
\rm{GeV}\lsim\mu\lsim10^{-18}\rm{GeV}$, the effect of moduli may be 
detectable by taking small $\ell$ into account. For $\mu$ smaller than 
$10^{-19}$GeV, $\sin^{2}2\theta_{M}$ is less than $10^{-2}$, which is 
so small that the conversion of $\nu_{\mu}$ to $\nu_{\tau}$ cannot be 
detected in long base line experiments. For $\mu$ lareger than 
$10^{-18}$GeV the effect is too small to discriminate it from 
oscillations in the vacuum. In case (A) of Fermilab, the range of 
$\mu$ where the effect may be observable shifts to a range around 
several times $10^{-18}$GeV. In case (B), $\sin^{2}2\theta_{M}$ is 
unity for any $\mu$ because of the maximal mixing angle $\theta=\pi/4$. 
The effect may only be seen in a small $\ell$. 

Next we turn to the repulsive force. The numerical results 
of $\sin^{2}2\theta_{M}$ and $\pi\ell^{-1}$ are listed in Table 3 and 4. 
In case (A) of Table 3, the moduli effect makes the values of $a$ large. 
Therefore, $\sin^{2}2\theta_{M}$ is so small that the effect is hard 
to observe. On the other hand, in case (B) of the CHORUS experiment, 
Table 3 shows that large $\phi$'s with small $\mu$'s 
($\lsim10^{-19}$GeV) enhance $\sin^{2}2\theta_{M}$ to be unity. 
On the long base line experiments (see Table 4), $\sin^{2}2\theta_{M}=1$ 
irrespective of the moduli effect. The effect may be seen only through 
the oscillation length.

We illustrate the oscillation probability as a function of the distance 
$L$[km] in case (A) for the CHORUS experiment with attractive force. 
In Fig. 2 we show the probability vs $L$ at near the resonance and 
the mass of the modulus is set at $\mu=3.94\times10^{-20}$[GeV]. The 
dotted line denotes the probability of the oscillation in the vacuum, 
and the solid line corresponds to the oscillation including moduli 
effect. The former magnitude, the value of which is 
$\sin^{2}2\theta_{M}=4\times10^{-4}$, 
is much smaller than the latter and changes much 
more frequently with $L$. 
The exact resonance occurs at $\mu=3.74\times10^{-20}$ [GeV] as shown 
in Fig. 3. 
The probability including the moduli effect increases more slowly than 
in Fig. 2 and reach the maximum value around $L=18$[km]. If such a bump 
is found experimentally, the mass of the modulus will be determined. 

It is noted that the values of $\mu$ in the discussion above must be 
changed if we take smaller values of $\alpha$ than the present ones 
which are upper limits in the experimental restrictions on the fifth 
force. However, as seen in Eq. (\ref{eqn:moduli}), a smaller $\mu$ 
coresponding to a smaller $\alpha$, which makes $\alpha\mu^{-2}$ 
invariant, gives a similar result on neglecting the $\alpha$ dependence 
on $\mu$. 

\vspace{1em}
{\large \bf 3.~Solar Neutrino Oscillations}\\

Now we will roughly examine to what degree the moduli interaction 
influences solar neutrino oscillations. We assume that $\nu_{e}$'s 
are generated in the region near the distance $R_{min}$ from the 
center of the Sun. While they propagate along the $R$ axis to the 
surface, they partly change into $\nu_{\mu}$. The final eigenstate 
of mass including the Mikheyev-Smirnov-Wolfenstein (MSW) effect 
\cite{MSW,MSW2} and also the moduli interaction is defined as 
$(\tilde{\nu_{1}}, \tilde{\nu_{2}})$ by which the flavor eigenstate 
$(\nu_{e}, \nu_{\mu})$ is written as 
\begin{equation}
\left(
\begin{array}{l}
\nu_{e} \\
\nu_{\mu}
\end{array}
\right)
=
\left(
\begin{array}{cc}
\cos\theta_{s}      & \sin\theta_{s}\\
-\sin\theta_{s}     & \cos\theta_{s}
\end{array}
\right)
\left(
\begin{array}{l}
\tilde{\nu_{1}} \\
\tilde{\nu_{2}}
\end{array}
\right),
\end{equation}
where $\theta_{s}$ is the sum of the mixing angles: One is from the 
eigenstate $(\nu_{1}^{s}, \nu_{2}^{s})$ of mass and the MSW effect to 
$(\nu_{e}, \nu_{\mu})$ and the other from ($\nu_{1}^{s}, \nu_{2}^{s}$) 
to $(\tilde{\nu_{1}}, \tilde{\nu_{2}})$. 

We use again a traceless Hamiltonian for $(\nu_{e}, \nu_{\mu})$:
\begin{equation}
i\frac{d}{dR}
\left(
\begin{array}{l}
\nu_{e} \\
\nu_{\mu}
\end{array}
\right)
=
\left(
\begin{array}{cc}
a_{s}(R)    & b_{s}(R)\\
b_{s}(R)    & -a_{s}(R)
\end{array}
\right)
\left(
\begin{array}{l}
\nu_{e} \\
\nu_{\mu}
\end{array}
\right),
\label{eqn:traceless}
\end{equation}
where
\begin{eqnarray}
a_{s}(R) &=& \frac{\Delta m'^{2}}{4E}\cos 2\theta ' 
-\frac{\sqrt{2}}{2}G_{F}N_{e}(R)
-\frac{\Delta f_{s}}{2}\phi (R), 
\label{eqn:cosine} \\
b_{s}(R) &=& \frac{\Delta m'^{2}}{4E}\sin 2\theta ' 
-f_{12}\phi (R).
\label{eqn:sine}
\end{eqnarray}
Here $\theta '$ is the mixing angle from the mass eigenstate with 
eigenvalues $m_{1}$ and $m_{2}$ to the flavor state, 
$\Delta m'^{2}=m_{2}^{2}-m_{1}^{2}$, and $\Delta f_{s}=f_{22}-f_{11}$.
In Eq. (\ref{eqn:cosine}), the term including Fermi's coupling constant 
$G_{F}$ and the number density of electrons, $N_{e}(R)$, represents 
MSW effect.
$N_{e}$ strongly depends on $R$ \cite{Bahcall}: 
\begin{equation}
N_{e}(R)=245N_{A}\exp\left(-10.54\frac{R}{R_{\odot}}\right)
\label{eqn:density}
\end{equation}
where $N_{A}$ is Avogadro's number and $R_{\odot}$ is the radius of the 
Sun. 

With respect to $\phi(R)$ in Eqs. (\ref{eqn:cosine}) and 
(\ref{eqn:sine}), assuming $\lambda=\mu^{-1}\ll R_{\odot}$, we can use 
Eq. (\ref{eqn:local}). By replacing $\rho_{sur}$ with $\rho_{\odot}$, 
we get
\begin{equation}
\phi(R) = \frac{4\pi G_{M}\rho_{\odot}(R)E}{\mu^{2}}
\left(1-\frac{2}{e}\right).
\end{equation}
Then the density of the Sun $\rho_{\odot}(R)$, is replaced using 
$N_{e}(R)$ as 
\begin{equation}
\rho_{\odot}(R)=\frac{m_{N}N_{e}(R)}{Y_{e}},
\end{equation}
where $m_{N}$ is the mass of a nucleon and $Y_{e}$ is the electron 
number per nucleon: $Y_{e}\simeq 1$. Eq. (\ref{eqn:traceless}) roughly 
leads to the probability at the distance $R$, 
similarly to Eq. (\ref{eqn:oscillation2})
\begin{equation}
P(\nu_{e}\rightarrow\nu_{\mu}, R)=
\frac{b_{s}(R)^{2}}{a_{s}(R)^{2}+b_{s}(R)^{2}}\sin^{2}
\left\{\int_{R_{min}}^{R} [a_{s}(R)^{2}+b_{s}(R)^{2}]^{1/2}dR\right\}. 
\label{eqn:oscillation3}
\end{equation}
The above equation reproduces the probability of MSW when $\phi(R)=0$.
We do not discuss this probability in detail, but examine the effect of 
moduli qualitatively. 

Let us consider the case (A$^{\scriptscriptstyle\prime}$), 
$\Delta f_{s}=1, f_{12}=0$, and case (B$^{\scriptscriptstyle\prime}$), 
$\Delta f_{s}=0, f_{12}=1$, seperately. In case 
(A$^{\scriptscriptstyle\prime}$), both $G_{F}$-term 
and $\phi$-term are proportional to $N_{e}(R)$. 
\begin{equation}
\left(\frac{\Delta f_{s}}{2}\right)\frac{\phi(R)}{N_{e}(R)} 
= 1.04\times 10^{-5}\left(\frac{\alpha}{10^{-4}}\right)
\left(\frac{\mu}{10^{-20}{\rm GeV}}\right)^{-2}
\left(\frac{E}{{\rm MeV}}\right)
\frac{1}{\rm{GeV}^{2}}.
\label{eqn:solar}
\end{equation}
Eq. (\ref{eqn:solar}) is equally matched with 
$(\sqrt{2}/2)G_{F}=8.25\times10^{-6}~~[\rm{GeV}^{-2}]$. 
Eq. (\ref{eqn:solar}) reads that when $\mu\sim 10^{-20}$GeV, 
$\alpha\sim 10^{-4}$, $E\sim 1$MeV, the $\phi$ term is comparable to 
the $G_{F}$ term in Eq. (\ref{eqn:cosine}) 
everywhere in the Sun. In addition to that, for 
$\Delta f_{s}\phi >0$ the resonance ($a_{s}=0$) occurs at a smaller 
value of $N_{e}(R)$ than that when only the MSW mechanism works. 
Resonance never occurs when $\Delta f_{s}\phi <0$ and 
$|\Delta f_{s}\phi|>(G_{F}$ term). 

In case (B$^{\scriptscriptstyle\prime}$) both terms of the 
right-hand side in Eq. (\ref{eqn:sine}) are reexpressed as 
\begin{equation}
\frac{\Delta m'^{2}}{4E}\sin 2\theta ' 
= 1.27 \times 10^{-3}\left(\frac{\Delta m'^{2}}{10^{-6}\rm{eV}^{2}}
\right)
\left(\frac{E}{{\rm MeV}}\right)^{-1} \sin 2\theta '
\frac{1}{\rm{km}},
\label{eqn:solvac}
\end{equation}
\begin{equation}
f_{12}\phi(R) = \phi(R)
= 4.90\times 10^{-4} \left(\frac{\alpha}{10^{-4}}\right)
\left(\frac{\mu}{10^{-20}{\rm GeV}}\right)^{-2}
\left(\frac{E}{{\rm MeV}}\right)
\left(\frac{\rho_{\odot}(R)}{{\rm g}{\rm cm}^{-3}}\right)
\frac{1}{\rm{km}}.
\label{eqn:solmod}
\end{equation}
Solar neutrino experiments suggest $\sin^{2}2\theta '\simeq 
3\times 10^{-2}$ and $\Delta m'^{2}\simeq {\rm several} \times 10^{-6}
{\rm eV}^{2}$. Then for $\rho_{\odot}(R)\simeq 1$ 
[g$\cdot\rm{cm}^{-3}$] and for other typical values expressed in Eqs. 
(\ref{eqn:solvac}) and (\ref{eqn:solmod}), both values are almost 
the same. This means that in the case of the attractive force 
[$f_{12}\phi(R)>0$] the probability is very small because of 
$b_{s}\simeq 0$. For $|\phi(R)|\gg\Delta m'^{2}/4E$, the probability 
is suppressed in case (A$^{\scriptscriptstyle\prime}$), but on the 
other hand increases in case (B$^{\scriptscriptstyle\prime}$). 

Next we will examine the argument of sine in Eq. (\ref{eqn:
oscillation3}) at the solar surface only in a simplified case. We 
compare the argument due to moduli terms alone with that only in the 
vacum. The integration with respect to $R$ is taken from $R_{min}=
0.1R_{\odot}$ to $R_{\odot}$. The vacuum part in the argument can 
be obtained easily from Eq. (\ref{eqn:solvac}) as
\begin{equation}
0.9\frac{\Delta m'^{2}}{4E}R_{\odot}=7.94\times 10^{2}
\left(\frac{\Delta m'^{2}}{10^{-6}{\rm eV}^{2}}\right)
\left(\frac{E}{{\rm{MeV}}}\right)^{-1}
\label{eqn:solvac2}
\end{equation}
On the moduli part, using Eq. (\ref{eqn:density}), we get 
\begin{equation}
\frac{1}{2}\int_{R_{min}}^{R_{\odot}}\phi(R)dR = 1.38\times10^{3}
\left(\frac{\alpha}{10^{-4}}\right)
\left(\frac{\mu}{10^{-20}{\rm GeV}}\right)^{-2}
\left(\frac{E}{{\rm MeV}}\right)
\label{eqn:solmod2}
\end{equation}
Comparing both values in Eqs. (\ref{eqn:solvac2}) and 
(\ref{eqn:solmod2}), moduli effect matches the oscillation in the 
vacuum if moduli mass is around $10^{-20}$GeV.

\vspace{1em}
{\large \bf 4.~Concluding Remarks} \\

In this paper we have pointed out new signals which would support the 
heterotic superstring theory in neutrino oscillation experiments. 
The theory always accomodates moduli. There are, however, too many 
candidates of the vaccum to determine the masses and the interactions of 
moduli. Here we assume that at least one modulus has a tiny mass such 
as $\mu\sim10^{-22}-10^{-14}$GeV. Its interaction is expected to depend 
on flavors and to affect base line and solar neutrino oscillations. 

The oscillation probability of $\nu_{\mu}$-$\nu_{\tau}$ in planning 
base line experiments and of $\nu_{e}$-$\nu_{\mu}$ in the Sun are 
numerically estimated. It is concluded that the effect of moduli is 
significant when its mass is less than or equal to $10^{-19}$GeV 
under our choice of the values of parameters. Note that when the 
mixing angle $\theta$ of the mass eigenstate to the flavor one is 
very small, the maximum value of the oscillation probability is very 
small in the vacuum. In CHORUS experiments, the value is $4\times10^{-4}$
(see Table 1). However, taking the moduli mass into account, a 
particular value of $\mu$ makes the maximum value of the probability 
unity as in the situation of the solar neutrino oscillation when the 
MSW mechanism exists. 

One of parameters is the ratio $\alpha$ of the moduli coupling constant 
to the gravitational constant. We took the values of $\alpha$ as 
maximum values satisfing restrictions from experiments on the fifth 
force. Such values of $\alpha$ are not ensured. As seen in 
Eq. (\ref{eqn:moduli}), however, by decreasing the value of $\mu$ for a 
smaller value of $\alpha$, we get a similar result. There are also 
ambiguities about the signs of the difference of coupling constants, 
$\Delta f(\Delta f_{s})$. 

Changing the neutrino energy $E$ and/or the length of the base line $L$, 
the effect of moduli varies. So if neutrino 
oscillation experiments are scrupulously performed with various 
conditions as well as solar ones, 
one may get a clue of the moduli to the form of interaction with matter 
and to its mass. In the present paper, we have given the basic 
equations for the oscillation probability with which one can estimate 
the moduli effect for a given condition. Note 
that though we have mentioned only moduli, the present results 
can be extended to other objects. Namely, any particle is a candidate 
which has a tiny mass and its interaction 
depends on flavors and if its strength is adequate. SUSY majoron 
\cite{majoron} might be one of them.
\vspace{1em}

{\large \bf Acknowledgments} \\

We would like to thank S. Aoki for useful comments.

\clearpage
Figure captions \\

Fig. 1: Schematic view of base line neutrino oscillations. The neutrino 
beam is injected from the accelerator at the point $x_{1}$ and detected 
at the point $x_{2}$. The base line length $L$ is the distance between 
$x_{1}$ and $x_{2}$.\\

Fig. 2: The $\nu_{\mu}$-$\nu_{\tau}$ oscillation probability as a 
function of the length. The solid line is drawn by taking numerical 
values in case (A) of Table 1 with $\mu=3.94\times 10^{-20}$[GeV]. 
The dotted line represents the probability in the vacuum. \\

Fig. 3: Same probability as in Fig. 2. The solid line means the same 
as in Fig. 2, but with $\mu=3.74\times 10^{-20}$[GeV] and shows a 
resonance behavior. The dotted line is the same as in Fig. 2.

\clearpage

\clearpage 

\begin{thebibliography}{1}
\bibitem{LEP}
 See, for example, LEP collaborations,
 Phys. Lett. {\bf B276}, 247(1992).

\bibitem{Gross}
 D. Gross, J. Harvey, E. Martinec and R. Rohm, 
 Nucl. Phys. {\bf B256}, 253(1985); 
 {\bf B267},75(1986).

\bibitem{Green}
 M. Green, J. Schwarz and E. Witten, {\it Superstring Theory 2} 
(Cambridge University Press, Cambridge, England, 1987); 
 M. Kaku, {\it Strings, Conformal Fields and Topology} 
 (Springer, New York, 1991). 

\bibitem{Ferrara}
 For a review, see S. Ferrara and S. Theisen, in {\it Elementary 
 Particle Physics}, Proceeding of the Third Hellenic Summer 
 School, Corfu, Greece, 1989, edited by E. N. Argyres {\it et al.} 
 (World Scientific, Singapore, 1990), pp. 620-656. 

\bibitem{Hyokyo}
 For example, see 
 A. Font, L. E. Ib\'{a}\~{n}ez, F. Quevedo and A. Sierra, 
 Nucl. Phys. {\bf B331}, 421(1990); 
 H. Kataoka, H. Munakata, H. Sato and S. Tanaka, 
 Phys. Lett. {\bf B289}, 321(1992).

\bibitem{Cvetic}
 M. Cveti\v{c}, Phys. Lett. {\bf B229}, 41(1989).

\bibitem{Macorra}
 A. de la Macorra, Phys. Lett. {\bf B335}, 35(1994).

\bibitem{Kelley}
 S. Kelley, J. L. Lopez, D. V. Nanopoulos and A. Zichichi, 
 Report No. CERN-TH. 7433/94; hep-ph/9409223, 1994(unpublished). 

\bibitem{Ibanez}
 L. E. Iba\~{n}ez and D. L\"{u}st, Phys. Lett., {\bf B267}, 51(1991). 

\bibitem{orbifold}
 For example, see L. Dixon, D. Friedan, E. Martinec and S. Shenker, 
 Nucl. Phys. {\bf B282}, 13(1987); 
 S. Hamidi and C. Vafa, Nucl. Phys. {\bf 279}, 465(1987); 
 A. Font, L. E. Iba\~{n}ez, F. Quevedo and A. Sierra, 
 Nucl. Phys. {\bf B331}, 421(1990).

\bibitem{Gasperini}
 M. Gasperini, Phys. Rev. {\bf D39}, 3606(1989) 

\bibitem{Nature}
 E. Fischbach and C. Talmadge, Nature (London) {\bf 356}, 207 (1992) 

\bibitem{Stacey}
 F. D. Stacey, G. J. Tuck and S. C. Holding {\it et al.}, Rev. Mod. 
 Phys. {\bf 59}, 157(1987).

\bibitem{Harari}
 H. Harari, Phys. Lett. {\bf B216}, 413(1989); Phys. Lett. {\bf B292}, 
 189(1992). 

\bibitem{Kamiokande}
 Kamiokande Collaboration, Y. Fukuda {\it et al.}, Phys. Lett. 
 {\bf B335}, 237(1994). 

\bibitem{CHORUS}
 CHORUS Collaboration, E. Arik {\it et al.}, Nucl. Inst. and Meth. 
 {\bf A360}, 254(1995). 

\bibitem{KEK}
 KEK proposal E-362 (unpublished). 

\bibitem{MINOS}
 MINOS Collaboration, Fermi-Lab p-875 (unpublished).

\bibitem{MSW}
 L. Wolfenstein, Phys. Rev. {\bf D17}, 2369(1978). 

\bibitem{MSW2}
 S. P. Mikheyev and A. Yu. Smirnov, Nuovo Cimento {\bf C9}, 17(1986). 

\bibitem{Bahcall}
 J. N. Bahcall and R. K. Ulrich, Rev. Mod. Phys. {\bf 60}, 297(1988). 

\bibitem{majoron}
 Y. Chikashige and T. Kon, report, hep-th/9510120, 1995 (unpublished); 
 J. Eliss, J. S. Hagelin, D. V. Nanopoulos and M. Srednicki,
 Nucl. Phys. {\bf B241}, 381(1984). 

\end{thebibliography}
\end{document}